\documentclass[iop,revtex4]{emulateapj}
\usepackage{graphics,graphicx}
\usepackage{helvet}
\usepackage[utf8]{inputenc}
\usepackage[T1]{fontenc}
\usepackage{longtable}
\usepackage{soul}
\usepackage{hyperref}
\usepackage{amsmath, amssymb, gensymb}
\usepackage{ulem}
\usepackage{natbib}
\usepackage{microtype}
\usepackage{multirow}
\usepackage{morefloats}
\usepackage{rotating}
\usepackage{xspace}
\usepackage{color}
\def\mjybm {mJy\,beam$^{-1}$\xspace}

\def\mjybmvert {$\left(\frac{\textrm{mJy}}{\textrm{beam}}\right)$\xspace}

\def\etal {\textit{et al.}\xspace}

\begin{document}


\slugcomment{\textit{Accepted for publication in ApJ on 22 Aug 2016}}

\title{880\,$\micron$ SMA polarization observations of the quasar 3C 286}
\shorttitle{SMA polarization observations of 3C 286}

\author{Charles~L.~H.~Hull\altaffilmark{1,2}} 
\author{Josep~M.~Girart\altaffilmark{1,3}} 
\author{Qizhou Zhang\altaffilmark{1}}

\shortauthors{Hull \etal}
\email{chat.hull@cfa.harvard.edu}

\altaffiltext{1}{Harvard-Smithsonian Center for Astrophysics, 60 Garden St., Cambridge, MA 02138, USA}
\altaffiltext{2}{Jansky Fellow of the National Radio Astronomy Observatory}
\altaffiltext{3}{Institut de Ci\`encies de l'Espai, (CSIC-IEEC), Campus UAB, Carrer de Can Magrans S/N, 08193 Cerdanyola del Vall\`es, Catalonia, Spain}

\begin{abstract}
\noindent
For decades the bright radio quasar 3C 286 has been widely recognized as one of the most reliable polarization calibrators at centimeter wavelengths because of its unchanging polarization position angle and high polarization percentage.  However, it has become clear in recent years that the polarization position angle of 3C 286 changes with increasing frequency, increasing from $\sim$\,33$\degree$ at $\lambda \gtrsim 3$\,cm to $\sim$\,38$\degree$ at $\lambda \approx 1$\,mm.  With the advent of high-sensitivity polarization observations by current and future (sub)millimeter telescopes, knowledge of the position angle of 3C 286 at higher frequencies is critical for calibration.  We report the first polarization observations of 3C 286 at submillimeter wavelengths, taken at 880\,$\micron$ (340\,GHz) with the Submillimeter Array.  We find a polarization position angle and percentage of $37.4 \pm 1.5\degree$ and $15.7 \pm 0.8$\%, respectively, consistent with previous measurements at 1\,mm. \\

\end{abstract}


\section{INTRODUCTION}
\label{sec:intro} 

3C 286 is a bright, highly polarized quasar at a redshift $z = 0.849$ \citep{Burbidge1969} with exceptionally stable polarization properties.  For decades 3C 286 has been known to have an unchanging polarization position angle of $33\degree$ (measured counter-clockwise, or east, from north) and a polarization percentage of $\sim$\,10\% at radio wavelengths ($\sim$\,1.4--6\,GHz; \citealt{Perley1982}), making it an ideal calibrator for radio telescopes in the northern hemisphere such as the Very Large Array (now the Karl G. Jansky Very Large Array; VLA) and the Very Long Baseline Array (VLBA).  \citet{Perley2013} performed dedicated VLA observations of 3C 286 and other calibrators from $\sim$\,1--44\,GHz, again confirming the robustness of 3C 286 as a polarization calibrator at $\nu \lesssim 6$\,GHz.  However, \citeauthor{Perley2013} note that the position angle of 3C 286 appears to increase slowly as a function of frequency above $\sim$\,9\,GHz, reaching an angle of $36\degree$ at 43.5\,GHz.  Higher frequency polarization observations of 3C 286 are critical if it is to be used as an absolute polarization calibrator at short wavelengths.

There have been several observations of 3C 286 at 3\,mm and 1\,mm that confirm this increase in polarization angle, including work by \citet{MarroneThesis, Agudo2012, Hull2015b, Nagai2016},\footnote{See also the Atacama Large Millimeter-submillimeter Array (ALMA) CASA Guide on 3C 286 polarization at 1\,mm; \citet{Nagai2016} report results from the same 1\,mm Science Verification data: \url{https://casaguides.nrao.edu/index.php/3C286_Polarization}} finding position angles of $\sim$\,36--41$\degree$, consistent with the increase seen by \citet{Perley2013}.  Here we report the first submillimeter polarization observations of 3C 286, taken at 880$\micron$ (340\,GHz) with the Submillimeter Array (SMA; \citealt{Ho2004}).\footnote{The Submillimeter Array is a joint project between the Smithsonian Astrophysical Observatory and the Academia Sinica Institute of Astronomy and Astrophysics, and is funded by the Smithsonian Institution and the Academia Sinica.}  We find a polarization position angle and percentage of $37.4 \pm 1.5\degree$ and $15.7 \pm 0.8$\%, respectively, consistent with previous measurements at 1\,mm, suggesting that the polarization angle plateaus at $\sim$\,38$\degree$ at high frequencies.


\vspace{0.2in}
\section{OBSERVATIONS AND CALIBRATION}
\label{sec:obs}

We performed 880$\micron$ SMA continuum polarization observations of 3C 286 ($\alpha_{\textrm{J2000}} =$ 13:31:08.2879, $\delta_{\textrm{J2000}} =$ +30:30:32.958) as a filler track (2015B-S068; PI: Charles L. H. Hull) on 2016 January 28 in the compact configuration, with a resolution of $\sim$\,$1\farcs8$. 

The observations were made using the SMA polarization receiver system \citep{MarroneThesis, Marrone2008b} in dual-receiver, full-polarization mode.  The lower sideband (LSB) and upper sideband (USB) of each receiver had 2\,GHz of bandwidth, ranging from 334.126--336.004\,GHz (LSB) and 344.126--346.004\,GHz (USB).\footnote{While the observations we present have a total bandwidth of 4\,GHz, the advent of the new SWARM correlator (SMA Wideband Astronomical ROACH2 Machine) will enable future dual-receiver, full-polarization observations with greater bandwidths.}  The average (band-center) frequencies of the two sidebands are 335.065\,GHz (LSB) and 345.056\,GHz (USB), corresponding to approximately 895 and 870\,$\micron$, respectively.

Unlike in single-receiver mode where wave-plate switching is required to measure all four cross correlations ($RR,LL,LR,RL$) between the left- ($L$) and right- ($R$) circularly polarized signals, in dual-receiver mode all correlations are measured simultaneously, with the 300\,GHz receiver measuring $L$ and the 400\,GHz receiver measuring $R$ when the quarter-wave plates are in the ``left'' position.  The quarter-wave plates on antennas 1 and 4 were periodically switched to the ``right'' position during observations of the bright calibrator 3C 84 in order to calibrate the cross-receiver delay and the $RL$-phase offset between the 300 and 400\,GHz receivers; this allowed us to calibrate the absolute position angle of the polarized radiation.

We performed the initial calibration---flagging, bandpass, cross-receiver delays, complex gains, flux calibration, and $RL$-phase offset---using the IDL superset \texttt{MIR} (see \citealt{QiYoung2015} for a description of how to calibrate polarization data in \texttt{MIR}).  Antennas 5 and 7 were out of the array.    The weather was excellent, with a 225\,GHz zenith opacity $\tau_{225} < 0.04$ for nearly the entire observation.  The double-sideband (DSB) system temperatures were between 145 and 216\,K.  

After initial calibration we exported the data to \texttt{MIRIAD} \citep{Sault1995}, which we used for measuring the polarization leakage terms and for imaging.  3C 286 has a flux of $\sim$\,240\,\mjybm at 880$\micron$, and was thus too weak to use as a polarization leakage calibrator.  Consequently, the 3C 286 observation was performed directly after another full-polarization track (2015B-S012, PI: Laurence Sabin), which included 3C 84, a bright calibrator suitable for polarization calibration.  After one round each of amplitude and phase self-calibration on 3C 84, we measured leakage terms and applied them to the 3C 286 data.  The leakage terms had amplitudes <\,1.5\% for all antennas in both sidebands.

\vspace{0.15in}
\section{RESULTS}
\label{sec:results}

We used \texttt{MIRIAD} to produce Stokes $I$, $Q$, and $U$ continuum images using natural weighting (\texttt{robust}\,=\,2, also \texttt{sup}\,=\,0 in \texttt{MIRIAD}).  The phases of 3C 286 were corrected with one round of phase-only self calibration.  The final synthesized beam is $1\farcs89\times1\farcs80$ at a position angle of $-19\degree$, and the rms noise levels in the Stokes $I,$ $Q,$ and $U$ maps are all $\sim$\,2\,\mjybm. 

\begin{figure} [hbt!]
\centering
\includegraphics[scale=0.35, clip, trim=0cm 0cm 0cm 0cm]{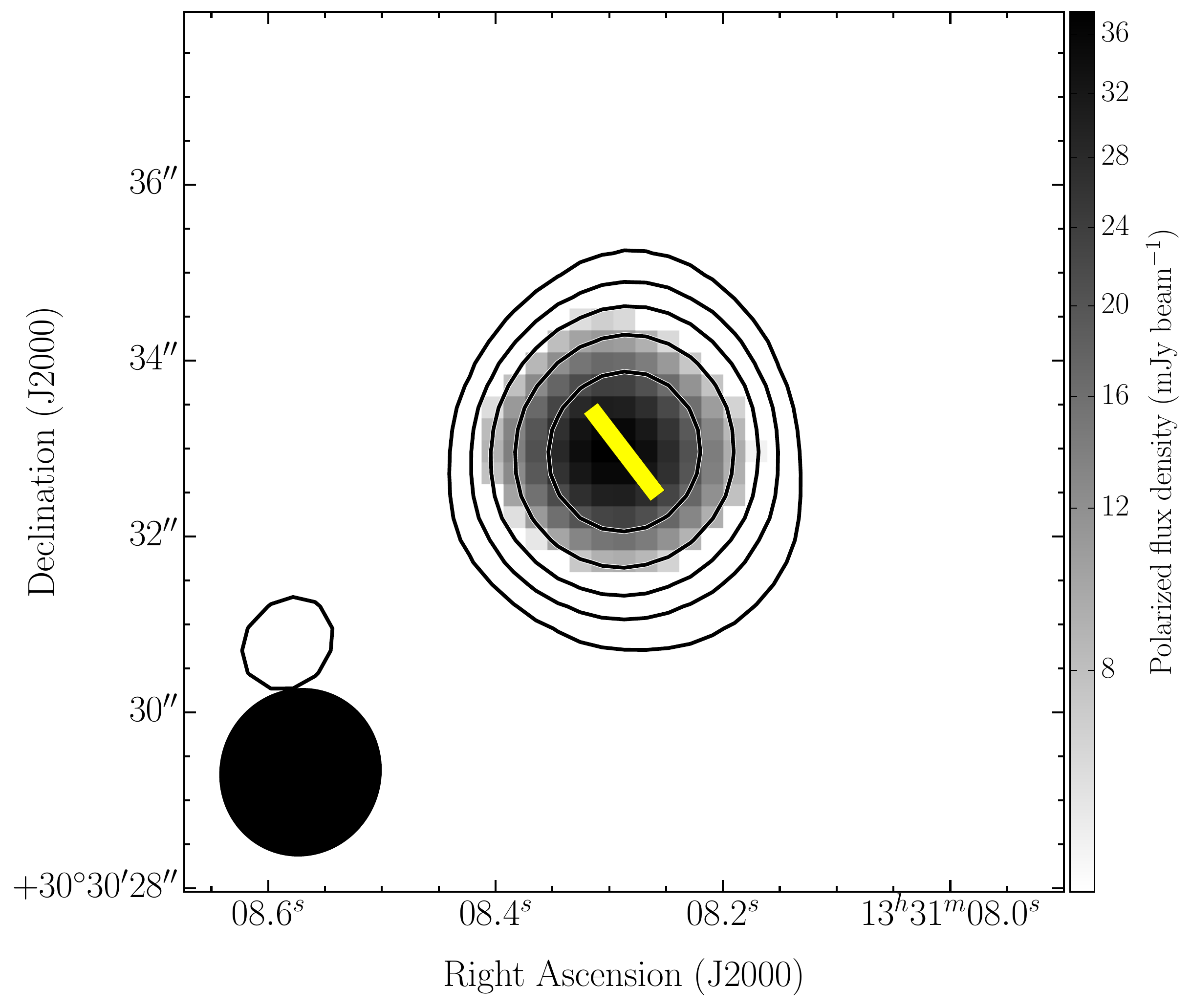}
\caption{\footnotesize   
880\,$\micron$ SMA image of 3C 286 ($\alpha_{\textrm{J2000}} =$~13:31:08.2879, $\delta_{\textrm{J2000}} =$~+30:30:32.958).  The rms noise level in the Stokes $I$ map $\sigma_I = 2$\,\mjybm.  The black contours are the Stokes $I$ total intensity at --3,\,3,\,8,\,16,\,32,\,64\,$\times$\,$\sigma_I$.  Grayscale is the polarized flux density $P = \sqrt{Q^2 + U^2}$.  The peak polarization percentage $P/I$ is $15.7 \pm 0.8$\%.  The yellow line segment is the polarization position angle $\chi = 37.4 \pm 1.5\degree$.  The ellipse in the lower left represents the synthesized beam, which measures $1\farcs89\times1\farcs80$ at a position angle of $-19\degree$.  
} 
\vspace{0.2in}
\label{fig:3C286_pol} 
\end{figure}

\begin{figure*} [hbt!]
\centering
\includegraphics[scale=0.35, clip, trim=0cm 0cm 0cm 0cm]{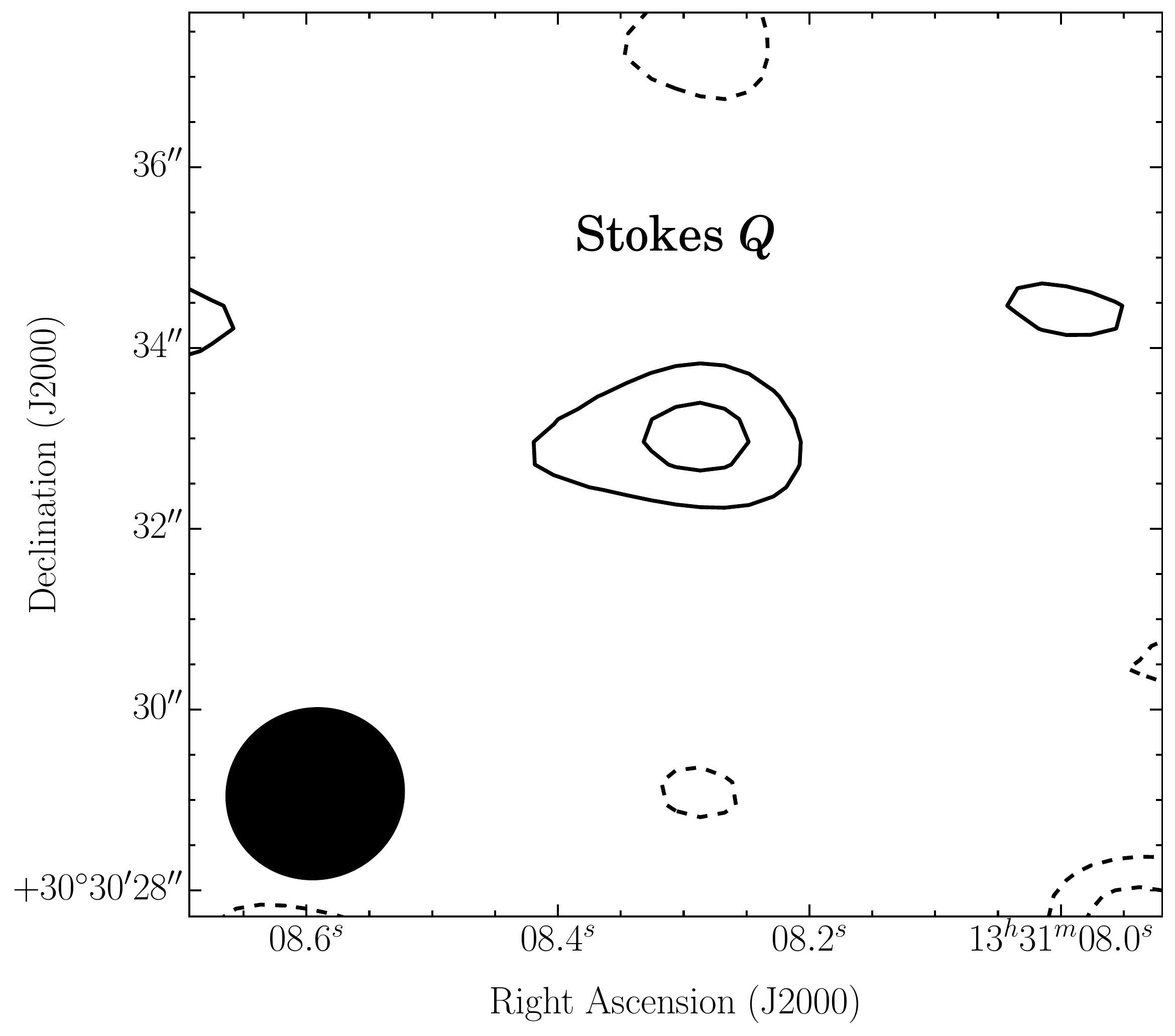}
\includegraphics[scale=0.35, clip, trim=0cm 0cm 0cm 0cm]{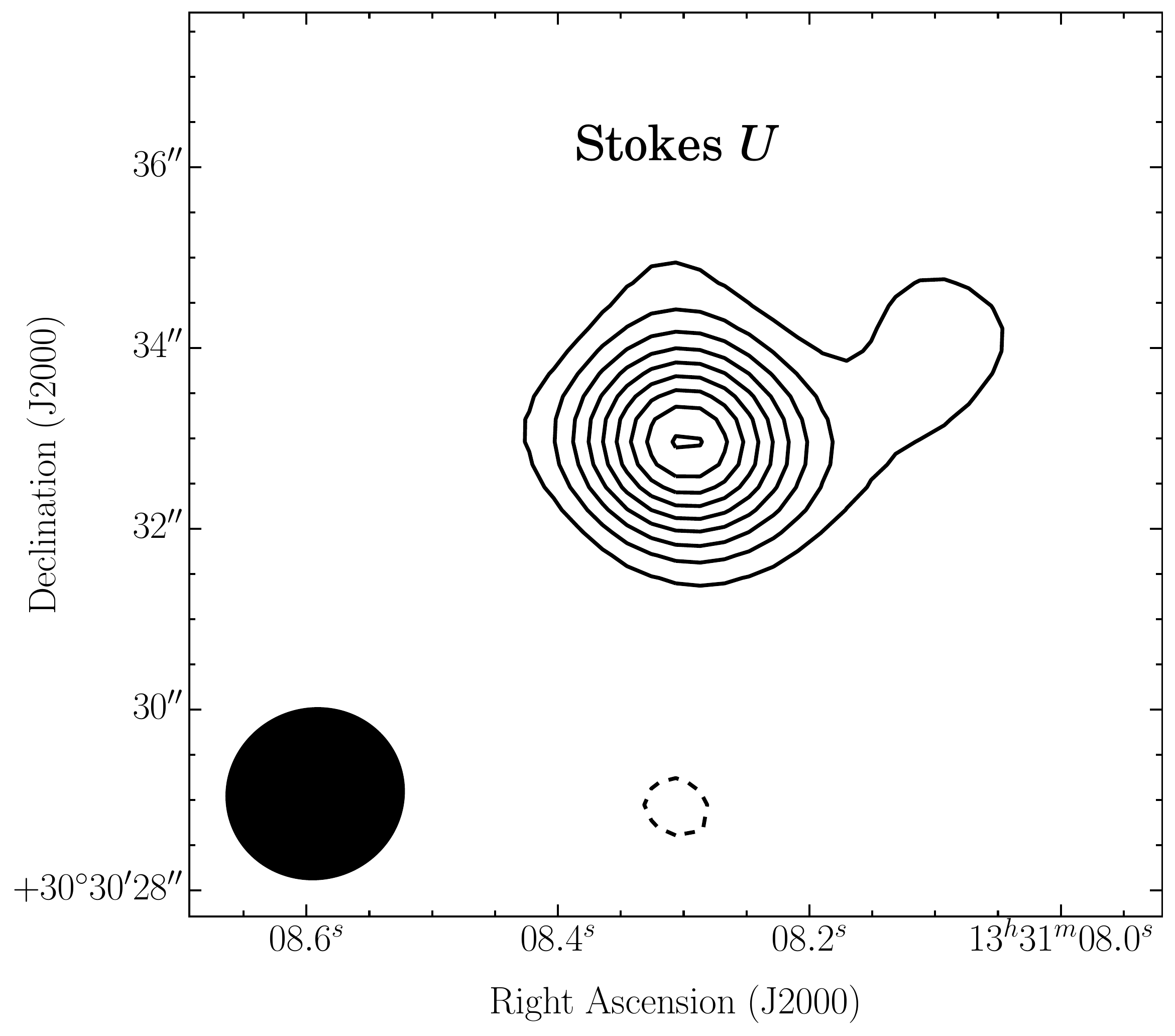}
\caption{\footnotesize   
Stokes $Q$ and $U$ maps of 3C 286.  The rms noise levels in the Stokes $Q,$ and $U$ maps $\sigma_Q = \sigma_U = \sigma_P = 2$\,\mjybm.  Black contours are the Stokes $Q$ (left) and $U$ (right) intensities at --3,\,--2,\,2,\,4,\,6,\,8,\,10,\,12,\,14,\,16,\,18\,$\times$\,$\sigma_P$.  As in Figure \ref{fig:3C286_pol}, the ellipses in the lower left represent the synthesized beam.
} 
\vspace{0.2in}
\label{fig:3C286_QU} 
\end{figure*}

We find that 3C 286 has a polarization position angle and percentage at 880\,$\micron$ of $37.4 \pm 1.5 \degree$ and $15.7 \pm 0.8\%$, respectively.  These results are consistent with previous measurements at 1\,mm.  We show the Stokes $I$ total intensity map with polarization orientation overlaid in Figure \ref{fig:3C286_pol}, and the Stokes $Q$ and $U$ maps in Figure \ref{fig:3C286_QU}.  Note that measurements of polarized intensity $P = \sqrt{Q^2 + U^2}$ should generally be debiased \citep{Vaillancourt2006, Hull2015b}; however, the signal-to-noise ratio (SNR) of $P$ always exceed $\sim$\,13, and thus the debiased value of the polarized intensity is unchanged to within the uncertainty of the original value.

Previous studies of 3C 286 with the VLA and CARMA (the Combined Array for Research in Millimeter-wave Astronomy) used the polarization of Mars to calibrate the absolute position angle of the telescope \citep{Perley2013, Hull2015b}, as rocky planets or satellites are expected to have polarization that is radial with respect to the planet's disk \citep{Heiles1963, Davies1966, White1973, Perley2013}.  Unfortunately, this test of absolute position angle has not yet been performed with the SMA dual-receiver polarization system.  However, the SMA has the unique advantage of using quarter-wave plates, which have very well known polarization properties \citep{MarroneThesis}.  Given a high enough SNR of the calibrator used to derive the cross-receiver delay and $RL$-phase offset, the polarization position angle accuracy is limited only by the rotation of the wave plates, which are positioned with an accuracy $\ll 1\degree$ \citep{MarroneThesis, Marrone2007}.



At (sub)millimeter wavelengths, 3C 286 has a polarization position angle $\chi \approx 38\degree$, where $\chi = 0.5 \arctan\left(U/Q\right)$. Somewhat inconveniently, this value is close to $\chi = 45\degree$,  where, by definition, all of the signal is in Stokes $U$ and none is in Stokes $Q$.  The 38$\degree$ position angle implies that the Stokes $Q$ signal is 4 times weaker than the Stokes $U$ signal.  At a high enough noise level (i.e., a SNR of the polarization emission $\lesssim 8$) the Stokes $Q$ signal would be undetectable, leading us to derive an incorrect position angle $38\degree < \chi < 45\degree$.  The value would be $< 45\degree$ because when calculating $\chi$ the value for $Q$ would not be identically zero, but would be an upper limit dictated by the rms noise in the Stokes $Q$ map.


Furthermore, the  standard CLEAN process on independent Stokes $Q$ and $U$ visibilities fails to properly account for the complex vector nature of the linear polarization when the signal is weak \citep[e.g.,][]{Pratley2016}.  \citet{Pratley2016} described a method for \texttt{CLEANing} that avoids this bias in low-signal-to-noise polarization observations, particularly in extended sources.  Fortunately, the combined (DSB) SMA data we present here are sensitive enough that the Stokes $Q$ map has a SNR $\gtrsim$\,3\,$\sigma_{P}$, and 3C 286 is a point source at SMA resolution, therefore this type of novel imaging approach is not necessary.

In short, the only results to date with high-SNR Stokes $Q$ maps are the 1\,mm ALMA results from \citet{Nagai2016} and the 880\,$\micron$ SMA results presented here, both of which yield polarization position angles of $\chi \approx 38\degree$, suggesting that there is no continued increase in the position angle of 3C 286 at $\lambda \lesssim 1$\,mm ($\nu \gtrsim 230$\,GHz).

\vspace{0.2in}
\section{DISCUSSION AND SUMMARY}

\begin{figure*} [hbt!]
\centering
\includegraphics[scale=0.35, clip, trim=1cm 0cm 1.5cm 0cm]{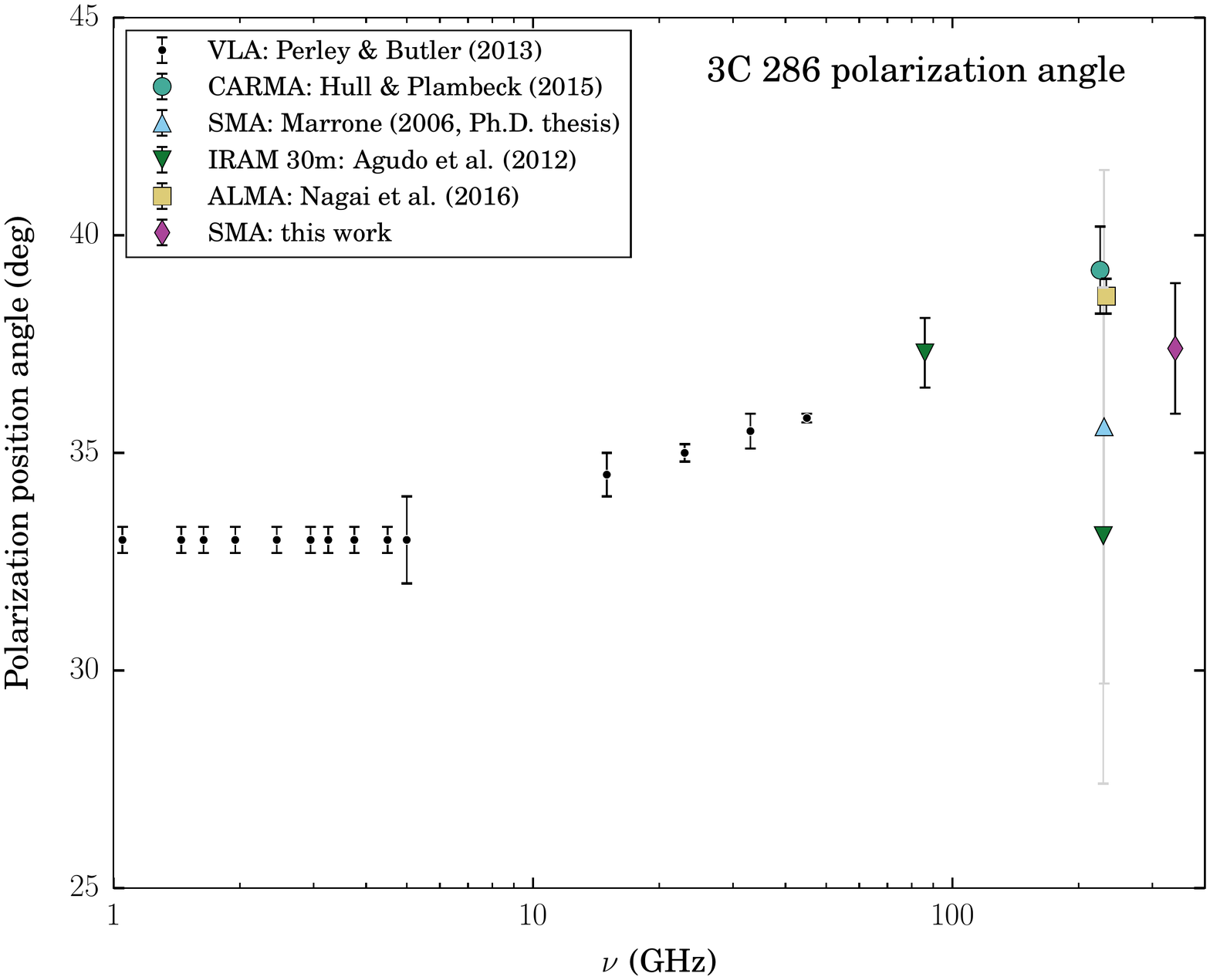}
\includegraphics[scale=0.35, clip, trim=1cm 0cm 2.5cm 0cm]{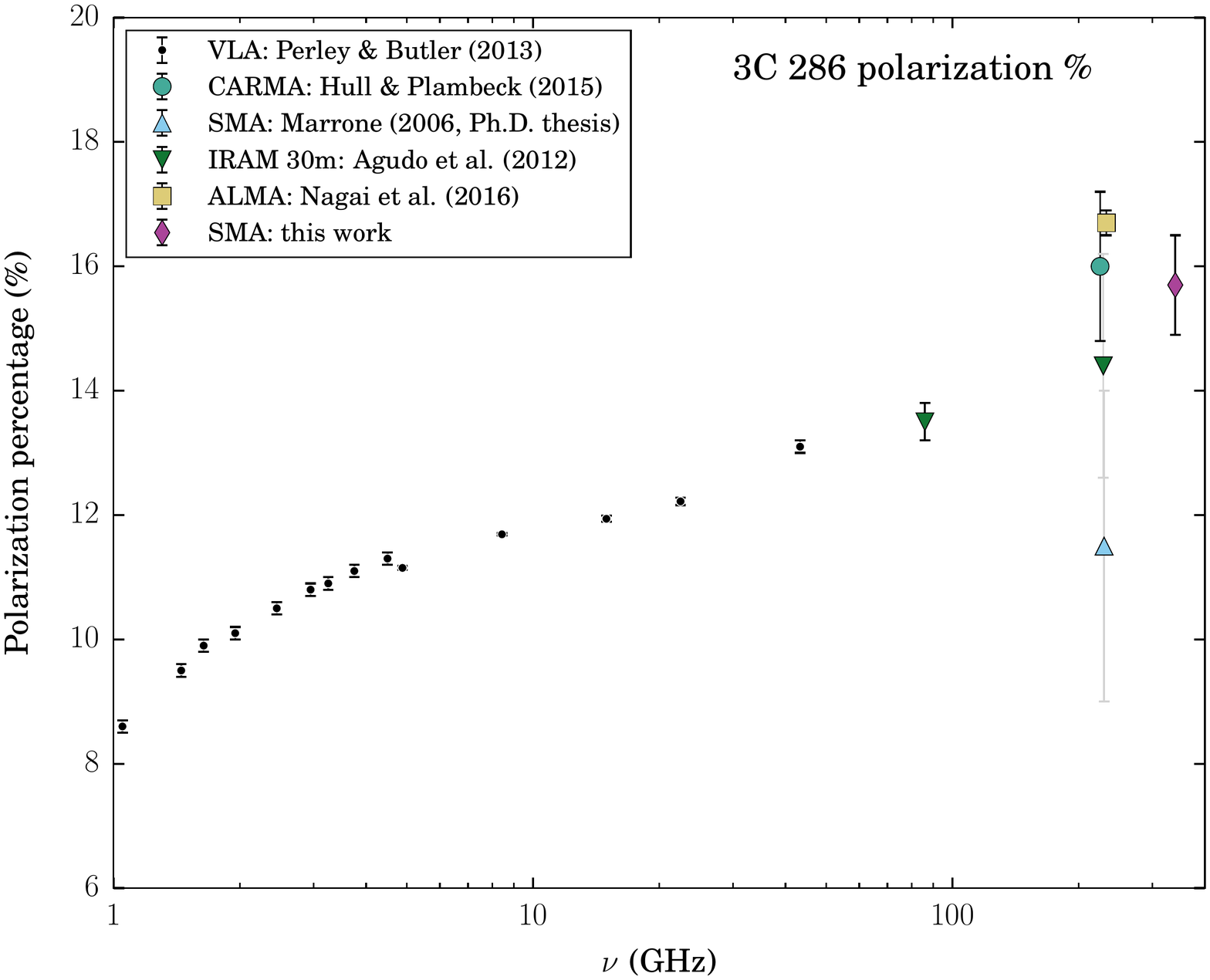}
\caption{\footnotesize   
3C 286 polarization position angle (left) and percentage (right) as a function of frequency.  
The data can be found in \citet{Perley2013} (VLA: $\sim$\,1--45\,GHz)$^\dagger$; \citet{Agudo2012} (IRAM 30\,m: 3\,mm, 1\,mm); \citet{Hull2015b} (CARMA: 1\,mm); \citet{Nagai2016} (ALMA: 1\,mm); \citet{MarroneThesis} (SMA: 1\,mm); and this work (SMA: 880$\micron$).
The CARMA polarization percentage was provided by Dick Plambeck (2016, private communication), and is associated with the 04 May 2014 observations listed in Table 3 of \citet{Hull2015b}.
The 1\,mm data points from \citet{Agudo2012} and \citet{MarroneThesis}, which have high uncertainties, have been plotted with light gray error bars to highlight the higher-SNR data. \\ \\
$^\dagger$ Polarization angle and percentage values at $\nu \leq 45$\,GHz include those from Table 3 ($\nu < 5$\,GHz) and Tables 2 and 5 ($\nu \geq 4.885$\,GHz) of \citet{Perley2013}.  The Table 3 values have been assigned errors in polarization percentage and angle of $\delta\Pi = 0.1\%$ and $\delta\chi = 0.3\degree$, where the former is the typical residual polarization percentage in a VLA map of an unpolarized source (Rick Perley, 2016, private communication), and the latter is calculated by standard error propagation: $\delta\chi = 0.5 \arctan{ \left(\delta\Pi / \Pi\right) }$.  Note that the $\nu > 4.885$\,GHz data from Tables 2 and 5 of \citealt{Perley2013} at are not at exactly the same frequency in the two plots.  The polarization percentage values at low frequencies have not been corrected for the increase as a function of time reported by \citet{Perley2013}.
} 
\vspace{0.2in}
\label{fig:3C286_PA_pct} 
\end{figure*}

In Figure \ref{fig:3C286_PA_pct} we extend the work of \citet{Perley2013} and show the polarization angle and percentage of 3C 286 as a function of frequency from $\sim$\,$1 - 340$\,GHz, now including the results from the Institut de Radioastronomie Millim\'etrique (IRAM) 30\,m telescope \citep[3\,mm, 1\,mm:][]{Agudo2012}, CARMA \citep[1\,mm:][]{Hull2015b}, ALMA \citep[1\,mm:][]{Nagai2016}, and the SMA (1\,mm: \citealt{MarroneThesis}; 880$\micron$: this work).

\begin{table}[tbh!]
\centerline{Table \ref{table:obs}: SMA 880\,$\micron$ polarization results  \vspace*{0.1in}}
\footnotesize
\begin{center}
\begin{tabular}{cccccc}
\hline \noalign {\smallskip}
$I$ & $Q$ & $U$ & $\chi$ & $\Pi$ \\
 \footnotesize \mjybmvert{} & \footnotesize \mjybmvert{} & \footnotesize \mjybmvert{} & ($\degree$) & (\%) \vspace{0.05in} \\
\hline \noalign {\smallskip}
$241.7 \pm 2.0$ & $10.0 \pm 2.0$ & $36.7 \pm 2.0$ & $37.4 \pm 1.5$ & $15.7 \pm 0.8$
\end{tabular}
\end{center}
\caption{\footnotesize 
Stokes $I$, $Q$, and $U$ fluxes, polarization position angle, and polarization percentage derived from the combined (DSB) 880\,$\micron$ (340\,GHz) SMA data.  The uncertainties in the $I$, $Q$, and $U$ maps are the rms noise values measured from the maps; the uncertainties in the polarization position angle and percentage are calculated using standard error propagation.}
\smallskip
\label{table:obs}
\end{table}


The polarization position angle of 3C 286 appears to be constant at $\sim$\,33$\degree$ at $\nu \lesssim$\,8\,GHz, and constant at $\sim$\,38$\degree$ at $\nu \gtrsim$\,230\,GHz.  The simplest explanation of the change in polarization position angle is a change in the position angle of either the jet or the magnetic field with distance from the nucleus, combined with a gradient in the spectral index of the emission along the jet.  In other words, we are most likely probing two different components at low and high frequencies: a population of less energetic electrons downstream in the jet that dominate the centimeter-wave emission, and a more energetic population of electrons further upstream (near the nucleus of the AGN) that dominate the (sub)millimeter-wave emission.  The observations at intermediate frequencies ($\sim$\,8--230\,GHz) where the polarization position angle changes from 33--38$\degree$ are probing a combination of the two regions.\footnote{The ``intermediate frequencies'' where the polarization position angle of 3C 286 changes could be anywhere from $\sim$8--86\,GHz to $\sim$\,8--230\,GHz, since there are no available polarization observations between the 86\,GHz observations from IRAM and the 230\,GHz observations from CARMA and ALMA.}

High-resolution (milliarcsecond-scale) observations of 3C 286 using very long baseline interferometry (VLBI) from $\sim$\,1.6--15\,GHz were performed in several studies including \citet{Zhang1994, Akujor1995, Jiang1996, Cotton1997, Nagai2016}, all of which report an elongated jet extending $\sim$\,50\,mas to the southwest of the nucleus.  \citet{Akujor1995, Jiang1996, Cotton1997} report polarization data, showing that both the nucleus and the SW jet have a polarization position angle that is consistent (on average) with the 33$\degree$ position angle assumed for 3C 286 at long wavelengths.  However, as noted above, the effective weighting of the polarization in the jet and the nucleus will be drastically different at (sub)millimeter frequencies.  Further discussion of the change in polarization position angle as a function of frequency can be found in \citet{Cotton1997, Agudo2012, Nagai2016}.

Because of the difficulty of high-frequency observations and the decreasing flux of 3C 286 at higher frequencies, polarimetric observations of 3C 286 at $\nu > 340$\,GHz ($\lambda$ < 880$\micron$) are impossible (or impractical) with current instruments.  To determine whether the position angle remains steady at $\sim$\,38$\degree$ at $\nu > 340$\,GHz will require confirmation by the ALMA Band 8 (440\,GHz), 9 (660\,GHz), and 10 (870\,GHz) polarization systems, which have yet to be commissioned.  Characterization of 3C 286 at these high frequencies will soon be essential: as millimeter, submillimeter, and far-infrared telescope sensitivity improves, 3C 286 will continue to be an invaluable calibrator for high-frequency polarization observations.


\acknowledgements
The authors thank the anonymous referee for the insightful comments.
C.L.H.H. thanks the SMA staff who made the SMA observations possible.  He particularly thanks Charlie Qi for his help with \texttt{MIR} data reduction.  He also thanks Ram Rao and Dan Marrone for the useful discussion of absolute polarization angle calibration of the SMA.
C.L.H.H. thanks Laurence Sabin for kindly allowing us to use data from her project 2015B-S012 to measure polarization leakage terms necessary to calibrate the 3C 286 data.
C.L.H.H. acknowledges Dick Plambeck for his help in deriving polarization percentages and uncertainties from the 1\,mm CARMA data, and Rick Perley for the discussion of the low-frequency data from \citet{Perley2013}.
J.M.G. acknowledges support from MICINN AYA2014-57369-C3-P and the MECD PRX15/00435 grants (Spain).
Q.Z. and J.M.G. acknowledge the support of the SI CGPS award, ``Magnetic Fields and Massive Star Formation.''
The National Radio Astronomy Observatory is a facility of the National Science Foundation operated under cooperative agreement by Associated Universities, Inc. 
This research made use of APLpy, an open-source plotting package for Python hosted at \url{http://aplpy.github.com}. \\ \\ \\ \\ \\ \\ \\ \vspace{-0.04in}

\bibliography{ms}
\bibliographystyle{apj}

\end{document}